# Twin-domain formation in epitaxial triangular lattice delafossites

*Jong Mok Ok[1,2,†], Sangmoon Yoon[1,†], Andrew R. Lupini[3], Panchapakesan Ganesh[3], Amanda Huon[1], Matthew F. Chisholm[3], Ho Nyung Lee[1,*]*

[1]Materials Science and Technology Division, Oak Ridge National Laboratory, Oak Ridge, TN 37831, U.S.A.

[2]Department of Physics, Pusan National University, Busan 46241, Korea

[3]Center for Nanophase Materials Sciences, Oak Ridge National Laboratory, Oak Ridge, TN 37831, U.S.A.

ABSTRACT. Twin domains are often found as structural defects in symmetry mismatched epitaxial thin films. The delafossite $AB$O$_2$, which has a rhombohedral structure, is a good example that often forms twin domains. Although bulk metallic delafossites are known to be the most conducting oxides, the high conductivity is yet to be realized in thin film forms. Suppressed conductivity found in thin films is mainly caused by the formation of twin domains, and their boundaries can be a source of scattering centers for charge carriers. To overcome this challenge, the underlying mechanism for their formation must be understood, so that such defects can be controlled and eliminated. Here, we report the origin of structural twins formed in a CuCrO$_2$ delafossite thin film on a substrate with hexagonal or triangular symmetries. A robust heteroepitaxial relationship is found for the delafossite film with the substrate, and the surface termination turns out to be critical to determine and control the domain structure of epitaxial delafossites. Based on such discoveries, we also demonstrate a twin-free epitaxial thin films grown on high-miscut substrates. This finding provides an important synthesis strategy for growing single domain delafossite thin films and can be applied to other delafossites for epitaxial synthesis of high-quality thin films.

$AB$O$_2$ delafossites are a new class of quantum materials with fascinating physical properties[1], ranging from the high electrical conductivity[2] to the giant Rashba spin splitting[3]. There have been several attempts to grow metallic delafossite thin films[4-9]. The growth of high-quality delafossite thin films, however, still remains a challenge because of their low-symmetry rhombohedral structure that limits the choice of substrate materials. Moreover, the formation of twin domains is often observed when grown on symmetry-mismatched commercial substrates. In fact, all the previous studies with metallic delafossite thin films, including PdCoO$_2$, PtCoO$_2$, and PdCrO$_2$, reported the formation of twin domains[4-9]. Although the metallic delafossite films exhibit comparable conductivity with the bulk value at room temperature ($\sigma^{\text{bulk}}/\sigma^{\text{film}}$ (300 K) ~ 2 – 10), the low temperature resistivity shows a huge deviation from the bulk value ($\sigma^{\text{bulk}}/\sigma^{\text{film}}$ (5 K) ~ 10$^3$ – 10$^4$) even in phase-pure thin films[8]. The previous observations suggested possible impurity scattering sources, such as twin-domain boundaries[9, 10], which tend to dominate over the electron-phonon scattering at low temperature. The severe deterioration of the electrical conductivity at low temperatures obstructs access to interesting physical phenomena, including hydrodynamic flow of electrons[11] and anomalous Hall effects[12,13]. To achieve highly conducting, ultra-pure delafossite thin films, therefore, the prerequisite is precise understanding of the formation of twin domains.

Thus, we investigated $CuCrO_2$ thin films to find the underlying mechanism for twin-domain formation. Compared with Pd-based metallic delafossites, $CuCrO_2$ is better suited to investigate the mechanism of twin-domain formation for the following two reasons: First, high-quality $CuCrO_2$ epitaxial films can be readily grown without any impurity phases[14], whereas other Pd- or Pt-based metallic delafossites are hard to grow epitaxially[4-9]. Second, a $CuCrO_2$ thin film can be used as a good buffer layer for the growth of metallic delafossite without interfering with the electrical transport chacterization[14] due to its high resistivity and excellent optical transparency. Thus, twin-domain-engineered $CuCrO_2$ thin films could serve as an excellent base for the subsequent epitaxy of metallic delafossite thin films.

In this report, we investigate twin-domain formation using $CuCrO_2$ thin films grown by pulsed laser epitaxy (PLE). Scanning transmission electron microscopy (STEM) studies reveals that the first $CrO_2$ layer maintains the octahedral symmetry by sharing oxygen atoms with the $Al_2O_3$ substrates, which critically determines the heteroepitaxial relationship between the film and substrate. This finding indicates that the atomic structure of the hexagonal substrate surface is the key to achieving single-domain delafossite thin films. We further found that the deliberate control of the hexagonal substrate surface leads to realization of twin-domain free delafossite thin films. Our results provide a fundamental basis for the successful growth of highly conducting Pd- or Pt-based metallic delafossite thin films.

$CuCrO_2$ thin films were grown by PLE using a KrF excimer laser ($\lambda$ = 248 nm) and a sintered polycrystalline target as described elsewhere[14]. To obtain atomically flat surfaces, $Al_2O_3$(0001) and $SrTiO_3$(111) substrates were chemically etched in a buffered hydrofluoric (BHF) acid followed by thermal annealing. Al$_2$O$_3$(0001) substrates were annealed at 1100 °C for 1 hour and SrTiO$_3$(111) substrates were thermally treated at 950 °C for 3 hours. The epitaxial CuCrO$_2$ thin films were grown on the treated substrates under optimal growth conditions of $T$ = 650 °C, $P_{O2}$ = 10 mTorr[14]. After the growth, the samples were cooled to room temperature in $P_{O2}$ = 100 Torr. The crystal structure was characterized by X-ray diffraction (XRD) using a four-circle high-resolution x-ray diffractometer (X'Pert Pro, PANalytical; Cu $K\alpha_1$ radiation). The surface morphology measurements were made with atomic force microscopy (Veeco Dimension 3100). High-angle annular dark-field scanning transmission electron microscopy (HAADF-STEM) images were collected with an aberration-corrected NION UltraSTEM 200 operated at 200 kV using a 30 mrad convergence semi-angle. Multislice STEM image simulations were carried out using QSTEM code. *Ab-initio* density functional theory (DFT) calculations were performed using the Vienna ab initio simulation package (VASP) code. The Perdew–Burke–Ernzerhof plus $U$ (PBE+$U$) was used for the exchange-correlation functional, in which the double-counting interactions were corrected using the full localized limit (FLL). The values used for $U$ parameters were 5 and 3 eV for Cu and Cr, respectively.

The $AB$O$_2$ delafossite has a natural superlattice structure (rhombohedral, space group R$\bar{3}$m), which consists of alternating stacks of closed-packed $A$ triangular lattice and edge-shared $B$O$_6$ tilted-octahedral layers as shown in Fig. 1(a). The stacked $A$ and $B$ layers of delafossite have an in-plane triangular symmetry as displayed in Fig. 1(b), which requires the use of substrates with a triangular (or hexagonal) symmetry, such as Al$_2$O$_3$(0001) and SrTiO$_3$(111). However, epitaxial films of rhombohedral structured materials are known to suffer from huge amounts of structural defects[15-17]. A twin boundary is the most common planer defect because of the simultaneous nucleation of two different domains, especially for thin films grown on substrates that have dissimilar structures with the films. In general, two different types of twin domains denoted by $T_1$ and $T_2$, which are associated by a 0° or 180° rotation, are found in such films. As experimentally observed, therefore, it is difficult to avoid the formation of the twin domains in the epitaxial thin films of delafossites.

Since the orientation relationship between the thin film and substrate is determined at the initial stage of epitaxial growth, it is important to understand the atomic structure of the interface. Fig. 2(a) shows a high-resolution HAADF STEM image of the interface between a $CuCrO_2$ thin film and an $Al_2O_3$ substrate seen along the $[\bar{1}100]$ $Al_2O_3$ zone axis. The brightest and second brightest dots in the image of the $CuCrO_2$ thin film indicate Cu and Cr atomic columns, respectively. Since the HAADF STEM provides scattering intensity that is approximately proportional to the square of the atomic number, O atomic columns are not visible in the image of a $CuCrO_2$ thin film. The HAADF STEM image shows that a coherent interface is formed between the $CuCrO_2$ thin film and the $Al_2O_3$ substrate. The $CrO_2$ sublayer is initially nucleated on the $Al_2O_3$ (0001) substrate, followed by the Cu sublayer thereon forming $CuCrO_2$. Such a coherent interface directly demonstrates that the delafossite thin film is grown on the substrate with a well-defined heteroepitaxial relationship.

To systematically understand the heteroepitaxial relationship between the $CuCrO_2$ thin film and $Al_2O_3$ substrate, its atomic interface structure was further analyzed. Provided that the first $CrO_2$ sublayer maintains the octahedral structure of the bulk state, the potential heteroepitaxial relationship is reduced to only two possible types: the type-I interface, in which two thirds of the Cr atoms are fully connected with top-surface Al atoms by edge sharing and one third of them by edge and corner sharing (Fig. 2(b)), and the type-II interface, in which two thirds of the Cr atoms are connected with the substrate by face sharing and one third by corner sharing (Fig. 2(c)). As a consequence, in-plane coordinates of the Cr atoms are slightly off from those of the Al atoms in the type-I interface, while the Cr atoms are positioned directly on top of the Al atoms in the type-II interface. Fig. 2(a) shows that the Cr and Al atoms are not in the same atomic plane at the experimental HAADF STEM image of the interface, strongly suggesting that the $CuCrO_2$ thin film is nucleated on the $Al_2O_3$ (0001) substrate with the type-I epitaxial relationship. To confirm the atomic structure of the interface, multislice HAADF STEM simulations were performed using the experimental parameters. As shown in Figs. 2(b) and (c), the simulated HAADF STEM image of the type-I interface agrees well with the experimental HAADF STEM image shown in Fig. 2(a).

To compare the thermodynamic stability of the two interface structures, the formation energy of the corresponding nucleation layers was calculated using slab interfacial models, where a surface as well as an interface are included (see Supporting Fig. S1 online). Note that the calculations based on these slab models are suitable for studying the orientational relationship of nucleation layers, since both surface and interface play an important role at the initial stage of the growth. The slab models and computational details are described in the Supporting Information. Notably, the formation energy of the type I interface is 0.87 eV lower than that of the type II interface (Fig. 2(d)), which further supports our experimental observations on the heteroepitaxial relationship.

The STEM and DFT results both suggest that the $CuCrO_2$ thin films are grown on $Al_2O_3$ (0001) substrates with only one orientational relationship, which could therefore result in a single domain structure of an epitaxial thin film. However, the $Al_2O_3$ corundum structure shown in Fig. 3(a) has two different types of termination on the (0001) surface, as shown in Fig. 3(b). $Al_2O_3$ has a hexagonal close-packed (hcp) crystal structure ($R\bar{3}c$ space group) with lattice constants $a = 0.476$ nm and $c = 1.299$ nm[18]. The crystal structure along the $c$-axis consists of alternating stacks of two different layers, which is caused by empty cation sites. Their surface termination and step height (~0.21 nm), which is a period of 1 monolayer (ML), were confirmed by AFM topographic investigation. When 1-ML step-and-terrace features are achieved on Al$_2$O$_3$(0001), A and B layer-terminated surfaces, corresponding to 0° and 180° rotations, respectively, should appear alternatively. Therefore, even though a preferential epitaxial relationship, i.e., the type-I epitaxial relationship, exists, twin domains are unavoidably formed because of the presence of rotated surface terminations, as schematically shown in the bottom of Fig. 3(b). This analogy means that the twin domains of the CuCrO$_2$ thin film on Al$_2$O$_3$ (0001) are formed not simply by a random match of the triangular lattice of the film with the hexagonal lattice, but rather by the orientationally-preferred growth on differently terminated surfaces.

This mechanism of twin-domain formation in CuCrO$_2$ thin films grown on Al$_2$O$_3$ (0001) substrates is further supported by the following experimental evidence. Figure 3(c) shows XRD $2\theta$-$\theta$ scans for a CuCrO$_2$ film grown on a Al$_2$O$_3$(111) substrate. $2\theta$-$\theta$ scans reveal that our film is of high quality without any impurity. Figure 3(d) shows XRD $\phi$ scans of the Al$_2$O$_3$ 01$\bar{1}$2 and CuCrO$_2$ 01$\bar{1}$2 peaks. The three peaks from the Al$_2$O$_3$ 01$\bar{1}$2 reflection are clearly separated by 120° due to the threefold symmetry. On the other hand, the CuCrO$_2$ 01$\bar{1}$2 revealed six peaks with similar intensities. This finding clearly supports the existence of twin domains with almost an equal probability, caused by the alternating surface termination of sublayers in Al$_2$O$_3$ substrates. Thus, the deliberate control of surface termination on hexagonal substrates can be an effective means to control the formation probability of each twin-domain, ultimately leading to growing twin-domain free thin films.

The easiest way to achieve the control of the hexagonal substrate surface is utilizing the different substrates such as SrTiO$_3$(111). Compared to Al$_2$O$_3$, SrTiO$_3$(111) has both advantages and disadvantages. The advantages include the fact that the surface termination can be controlled by chemical etching[19]. There are two possible terminations of the SrTiO$_3$(111) surface[20]: SrO$_3^{4-}$ and Ti$^{4+}$. For SrTiO$_3$(111), the Ti-termination with an atomically flat surface can be achieved by etching in a BHF solution and thermal annealing[21, 22, 23] (grey colored triangles in Fig. 4(a) show Ti-terminated surfaces). The Ti-termination surface may help stabilize only one domain out of the two twin domains. However, the SrTiO$_3$(111) surface is polar, in which a surface reconstruction and formation of TiO$_x$ surface are naturally expected[20]. The reconstruction presumably makes the initial nucleation of the pure delafossite phase more complicated, and consequently the flat surface will not provide enough nucleation sites for the orientationally-preferred growth. In this case, the step terrace, which is determined by the miscut angle of substrates, serves as the important nucleation sites and can determine the heteroepitaxial relationship between the film and substrate[24]. To check the controllability of twin domains, therefore, we grew CuCrO$_2$ thin films on SrTiO$_3$(111) substrates with different miscuts.

Figure 4(b) shows XRD $2\theta$-$\theta$ scans for CuCrO$_2$ films grown on Ti-terminated SrTiO$_3$(111) substrates with two different miscut angles (0.05 and 0.2º). All the film peaks in $2\theta$–$\theta$ scans are CuCrO$_2$ 0003n peaks, which suggest that CuCrO$_2$ films grown on SrTiO$_3$(111) substrates are high-quality without any impurity phases. Figure 4(c) shows XRD $\phi$ scans for the SrTiO$_3$ 110 and CuCrO$_2$ 01$\bar{1}$2 peaks grown on SrTiO$_3$(111) substrates with different miscuts. The $\phi$ scans confirmed the threefold symmetry of the SrTiO$_3$(111) substrate, while the CuCrO$_2$ thin film grown on the low miscut SrTiO$_3$(111) substrate (~0.05º miscut) showed six peaks with a 60° interval. Importantly, one set of three peaks revealed much smaller intensities than the other, suggesting that one domain might be prevailingly formed on the SrTiO$_3$(111) surface. This result implies not only that the domain formation is closely related to the atomic structure of the substrate surface, but also that there is a possibility of engineering the twin-domain formation by controlling the substrate surface.

To check the viability for achieving single-domain films, we further investigated a CuCrO$_2$ film on the high miscut SrTiO$_3$(111) substrate (~0.2º miscut). Interestingly, an XRD $\phi$ scan of a CuCrO$_2$ film grown on the high miscut substrate clearly revealed that one of the twin domains is fully suppressed, forming a single-domain film as shown in Figure 4(c) (red curve). To further confirm the single-domain, pole figure measurements of for the CuCrO$_2$ $01\bar{1}2$ peak ($2\theta = 36.33º$) were performed. As seen in Figure 4(d), the pole figure shows only three peaks at $\chi = 73.3º$, confirming again the single-domain in the CuCrO$_2$ thin film. Compared to low miscut substrates, high miscut SrTiO$_3$(111) substrates generally have more step terraces, in which the dominant nucleation for orientationally-preferred growth occurs (see Fig. S3 in the Supporting Information). The miscut of substrates, therefore, plays an important role in guiding the crystallographic orientation, leading to the single-domain thin films.

Overall, our results provide valuable insights into the formation of twin domains and how to control them to grow twin-free delafossite thin films. We find that the formation of twin domains strongly depends on the atomic structure of the substrate surfaces, which provides a control knob for twin domains. We successfully achieve this control by using high-miscut substrates as demonstrated in this work. In addition to our approach, there might be other routes to realize single-domain delafossite thin films. One possible solution could be the use of a step bunched surface of Al$_2$O$_3$. It was reported that a nonpolar single terminated surface could be achieved in Al$_2$O$_3$ by using step bunching[25,26]. Another possible solution could be the use of insulating delafossite single crystals as substrates. Among these candidates, Cu-based delafossite crystals provide the best option for a substrate[27] because of its insulating and optically transparent properties.

In summary, we have shown a possibility of controlling twin domains in delafossite thin films. We have studied the interface of delafossite thin films because the formation of twin domains is mainly determined at the initial nucleation stage. We found that the first $CrO_2$ nucleation layers share oxygen atoms with the substrate to maintain the $CrO_6$ octahedral structure. Through this initial nucleation process, the film and substrate have a preferential heteroepitaxial relationship. By exploiting this heteroepitaxial relationship, we demonstrated a growth approach to significantly reduce the formation of twin domains and to ultimately achieve single-domain films, as we confirmed in films grown on high-miscut $SrTiO_3(111)$ substrates. Overall, our results could lead to growing domain-free metallic delafossite thin films that will enable utilization of the highly mobile electrons, with which intriguing physics and novel ballistic electronic devices might be realized.

## ASSOCIATED CONTENT

**Supporting Information**. The Supporting Information is available free of charge at ****.

Figure S1: Atomic structure of type-I and II nucleation slab model for DFT calculations.

Figure S2: Miscut steps of $SrTiO_3(111)$ substrates.

Figure S3: Supplemental HAADF STEM images of a $CuCrO_2$ thin film grown on an $Al_2O_3(0001)$ substrate.

Figure S4: Low- and high-magnification HAADF STEM images of a $CuCrO_2$ thin film grown on a $SrTiO_3(111)$ substrate.

## AUTHOR INFORMATION

**Corresponding Author**

*Correspondence should be addressed to hnlee@ornl.gov.

**Present Addresses**

**Author Contributions**

J.M.O., S.Y. and H.N.L. designed the experiment and wrote the manuscript with inputs from all authors. J.M.O. and A.H. grew films and conducted XRD measurements. S.Y., A.R.L. and M.F.C conducted STEM experiments. S.Y. and P.G. performed DFT calculations. All authors have given approval to the final version of the manuscript. ‡These authors contributed equally.

**Funding Sources**

This work was supported by the U.S. Department of Energy, Office of Science, Basic Energy Sciences, Materials Sciences and Engineering Division (synthesis and microscopy) and as part of the Computational Materials Sciences Program (theory). This research used resources of the National Energy Research Scientific Computing Center (NERSC), a U.S. Department of Energy Office of Science User Facility operated under Contract No. DE-AC02-05CH11231.

**Notes**

ACKNOWLEDGMENT

ABBREVIATIONS

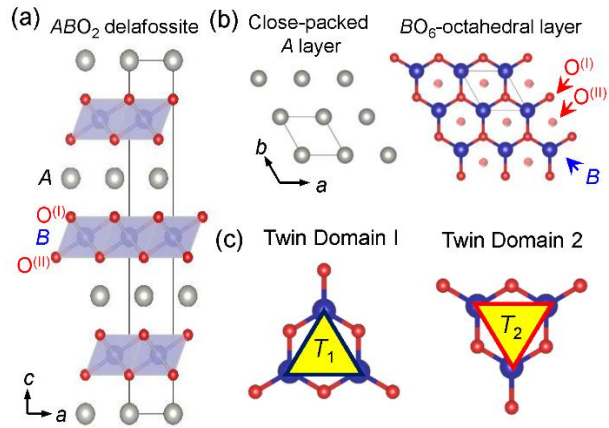

**Figure 1.** (a) Crystal structure of the delafossite $ABO_2$ (space group $R\bar{3}m$). $A$, $B$, and O atoms are colored in silver, blue, and red, respectively. (b) Top views of closed-packed $A$ and edge-sharing $BO_6$ octahedral layers (c) Schematics for the two delafossite domains.

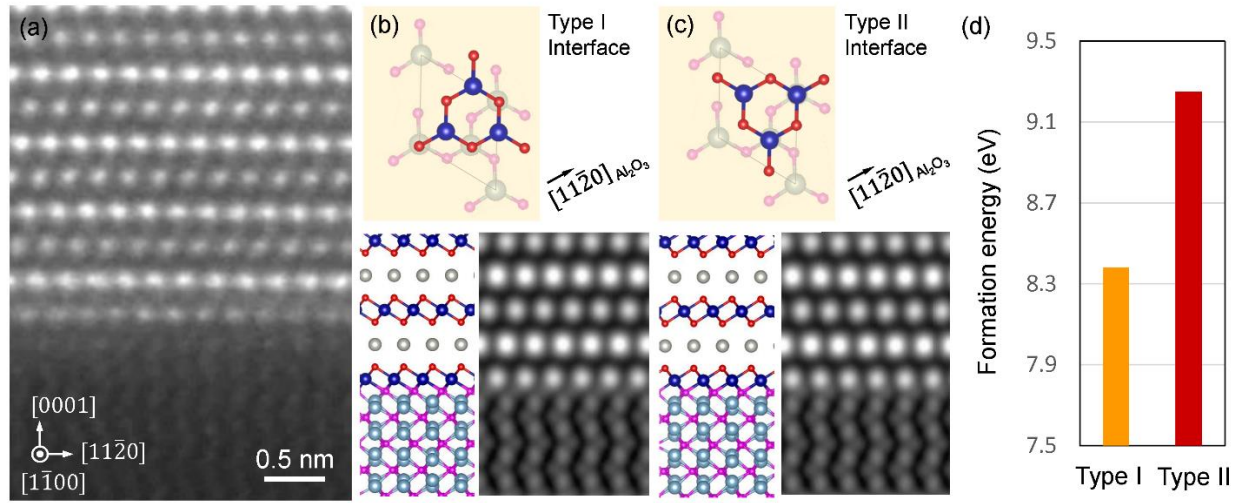

**Figure 2.** (a) HAADF STEM image of a $CuCrO_2$ thin film grown on an $Al_2O_3$ (0001) substrate seen along the [1$\bar{1}$00] zone axis. (b), (c) Schematics of interfacial atomic structures (top) and simulated HAADF STEM images of (b) the type I and (c) type II interfaces. Cross-sectional views of the interface atomic models (bottom left) and the corresponding simulated HAADF image (bottom left) are also shown. The O atoms of $CuCrO_2$ and $Al_2O_3$ are distinctively denoted by red and pink colors. (d) The formation energy for the type-I and type-II domains calculated by DFT.

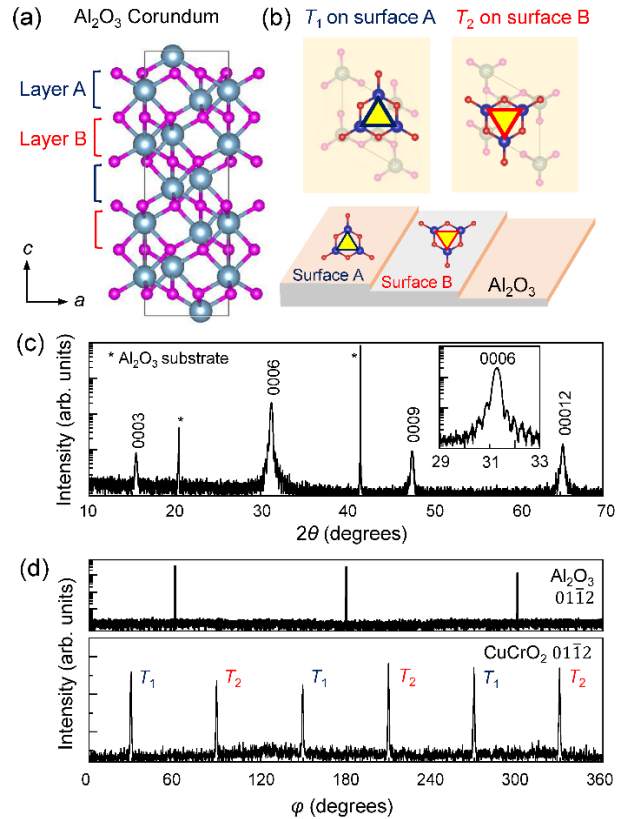

**Figure 3.** (a) Crystal structure of the corundum $Al_2O_3$ (space group $R\bar{3}c$), which consists of two stacking layers of A and B. (b) The twin domains matched on A and B surfaces of $Al_2O_3$. The schematic of twin-domain formation on step-and-terrace surfaces of a treated $Al_2O_3$ substrate is shown in the lower panel (b). (c) XRD $2\theta$–$\theta$ pattern of a $CuCrO_2$ thin film grown on an $Al_2O_3$(0001) substrate. (d) XRD in-plane azimuthal $\phi$ scans with the $01\bar{1}2$ reflection for an $Al_2O_3$ substrate (top) and a $CuCrO_2$ thin film (bottom).

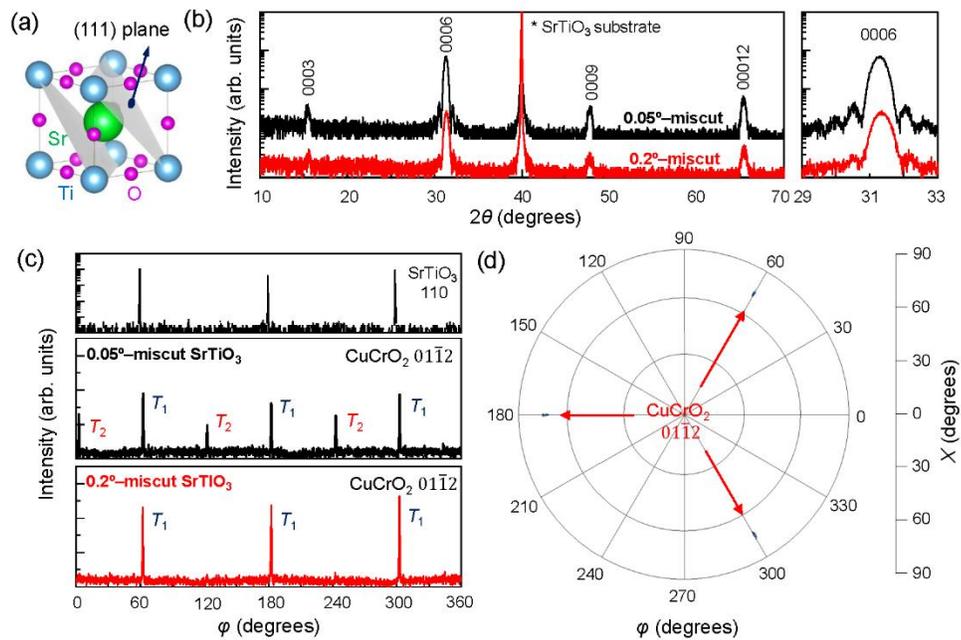

**Figure 4.** (a) Crystal structure of $SrTiO_3$. $SrTiO_3$(111) planes are colored in grey. (b) XRD $2\theta$-$\theta$ pattern of $CuCrO_2$ thin films on $SrTiO_3$(111) substrates with different miscut angles (black: 0.05º and red: 0.2º). (c) XRD $\phi$ scans for the $SrTiO_3$ substrate (top) and $CuCrO_2$ thin films grown on 0.05-miscut (middle) and 0.2º-miscut (bottom) $SrTiO_3$ substrates. (d) Pole figure of the $CuCrO_2$ $01\bar{1}2$ peak, confirming the single domain configuration with a three-fold symmetry.